\documentclass[aps,prl,reprint,superscriptaddress]{revtex4-1}
\usepackage[colorlinks,citecolor=blue]{hyperref}
\usepackage{textcomp}
\usepackage{subfigure}
\usepackage{verbatim}
\usepackage{bm}
\usepackage{hyperref}
\usepackage{amsmath}
\usepackage{graphicx}
\usepackage{epstopdf}
\usepackage{amssymb}
\usepackage{extarrows}
\usepackage{color}
\usepackage{CJK}
\usepackage{cancel}
\usepackage{ulem}
\usepackage[utf8]{inputenc}
\usepackage{url}
\newcommand{\ba}{\begin{eqnarray}}
\newcommand{\ea}{\end{eqnarray}}
\newcommand{\be}{\begin{equation}}
\newcommand{\ee}{\end{equation}}

\newcommand{\oct}{\mathrm{oct}}

\newcommand{\au}{\mathrm{AU}}

\newcommand{\IN}{\mathrm{in}}
\newcommand{\OUT}{\mathrm{out}}

\def\e1{e_1^2}

\definecolor{ochre}{rgb}{0.8, 0.47, 0.13}

\begin{document}
\title{Extreme resonant eccentricity excitation of stars around merging black-hole binary}
\author{Bin Liu}
\email{liubin23@zju.edu.cn}
\affiliation{Institute for Astronomy, School of Physics, Zhejiang University, 310058 Hangzhou, China}
\affiliation{Niels Bohr International Academy, Niels Bohr Institute, Blegdamsvej 17, 2100 Copenhagen, Denmark}
\author{Dong Lai}
\affiliation{Department of Astronomy, Center for Astrophysics and Planetary Science, Cornell University, Ithaca, NY 14853, USA}
\affiliation{Tsung-Dao Lee Institute, Shanghai Jiao Tong University, 200240 Shanghai, China}

\begin{abstract}
{We study the dynamics of a star orbiting a merging black-hole binary (BHB) in a coplanar triple configuration.
During the BHB's orbital decay, the system can be driven across the apsidal precession resonance,
where the apsidal precession rate of the stellar orbit matches that of the inner BHB.
As a result, the system gets captured into a state of resonance advection until the merger of the BHB,
leading to extreme eccentricity growth of the stellar orbit.
This resonance advection occurs when the inner binary has a non-zero eccentricity and unequal masses.
The resonant driving of the stellar eccentricity can significantly alter the hardening rate of the inner BHB,
and produce observational signatures to uncover the presence of nearby merging or merged BHBs.
}
\end{abstract}

\maketitle

\textit{Introduction.---}\label{sec 1}
About 90 double compact object mergers have been detected by the LIGO-Virgo-KAGRA Collaboration in recent years \cite{LIGO-2021}.
Stellar-mass black-hole binary (BHB) mergers can arise through diverse formation channels, including the standard isolated binary evolution
\cite{Lipunov-1997,Lipunov-2007,Podsiadlowski-2003,Belczynski-2010,Belczynski-2016,Dominik-2012,Dominik-2013,Dominik-2015,Alejandro-2017},
chemically homogeneous evolution \cite{Mandel-2016,Marchant-2016,duBuisson,Riley}, and multiple-body evolution in the gas disks of active galactic nuclei
\cite{Baruteau-2011,McKernan-2012,McKernan-2018,Bartos-2017,Stone-2017,Leigh-2018,Secunda-2019,Yang-2019,Grobner-2020,Ishibashi-2020,Tagawa-2020,
Liyaping-2021,Ford-2021,Samsing-Nature,Lirixin-2022,Lijiaru-2022}.
Additionally, various dynamical channels involve strong gravitational scatterings in dense clusters
\cite{Zwart(2000),OLeary(2006),Miller(2009),Banerjee(2010),Downing(2010),Ziosi(2014),Rodriguez(2015),Samsing(2017),
Samsing(2018),Rodriguez(2018),Gondan(2018)}, tertiary-induced mergers
via von Zeipel-Lidov-Kozai (ZLK) oscillations
\cite{vonZeipel,Lidov,Kozai,Smadar,Miller-2002,Wen-2003,Antonini-2012,Antonini(2017),Silsbee(2017),Petrovich-2017,Liu-ApJ,Xianyu-2018,
Hoang-2018,Liu-Quadruple,Fragione-Quadruple,Fragione-nulearcluster,Zevin-2019,Liu-HierarchicalMerger}
or flyby-induced mergers \cite{Michaely-2019,Michaely-2020}.

In the standard ZLK-induced merger scenario, the tertiary companion must be in a sufficiently
inclined orbit to induce eccentricity excitation in the inner binary. However,
an alternative eccentricity-pumping mechanism, known as apsidal procession resonance,
has been proposed for triples with small mutual inclinations (with no ZLK oscillations) \cite{Liu-Yuan,Liu-2020-PRD}.
In this mechanism, the apsidal precession rates of both the inner and outer orbits are in a 1:1 commensurability i.e.,
\be\label{eq:APR}
\dot\varpi_\IN\simeq\dot\varpi_\OUT,
\ee
allowing efficient eccentricity transfer between the two orbits.
This resonance, driven by the lowest-order post-Newtonian (PN) effect for the inner binary ($\dot\varpi_\IN$)
and the Newtonian interaction between the inner and outer orbits ($\dot\varpi_\OUT$),
can produce merger events with either low-mass or supermassive black-hole tertiary companions \cite{Liu-Yuan,Liu-2020-PRD}.

BHB progenitors, potentially numerous yet undetected, could form an abundant population in stellar multiples
(triples or quadruples) \cite{Tokovinin,Raghavan,Fuhrmann,Moe-2017}.
Observationally, triple systems with massive stars are suggested to be less inclined or nearly coplanar,
especially if the tertiary is not so faraway \cite{Ransom,Thompson,Eisner}.
If the tertiary companion is a visible object (e.g., a star or a pulsar),
its motion can aid in the search for inspiral BHBs \cite{Suto-1,Suto-2,Suto-3,Liu-Daniel}.
In this \textit{letter}, we focus on the secular dynamics of a star orbiting a merging BHB in a coplanar setup.
Beginning with an eccentric BHB and a nearly circular outer stellar orbit,
we demonstrate that the system can enter an apsidal resonance and be captured into a `resonance advection' state
during the inner BHB's gravitational radiation driven orbital decay.
Once captured, the eccentricity of the star grows continually,
attaining a very large value ($\gtrsim 0.9$).

\textit{Orbital evolution.---}\label{sec 2}
Consider an inner BHB with masses $m_1$, $m_2$, and
a tertiary companion ($m_3$) orbiting the center of mass of the inner bodies.
The reduced mass for the inner binary is $\mu_\IN\equiv m_1m_2/m_{12}$ where $m_{12}\equiv m_1+m_2$.
Similarly, the outer binary has $\mu_\OUT\equiv(m_{12}m_3)/(m_{12}+m_3)$.
The semi-major axes and eccentricities are denoted by $a_\IN$, $a_\OUT$ and $e_\IN$, $e_\OUT$, respectively.
Therefore, the orbital angular momenta of the two orbits are given by
$L_\IN=\mu_\IN\sqrt{G m_{12}a_\IN(1-e_\IN^2)}$
and $L_\OUT=\mu_\OUT\sqrt{G m_{123}a_\OUT(1-e_\OUT^2)}$.

\begin{figure}
\begin{centering}
\includegraphics[width=8.8cm]{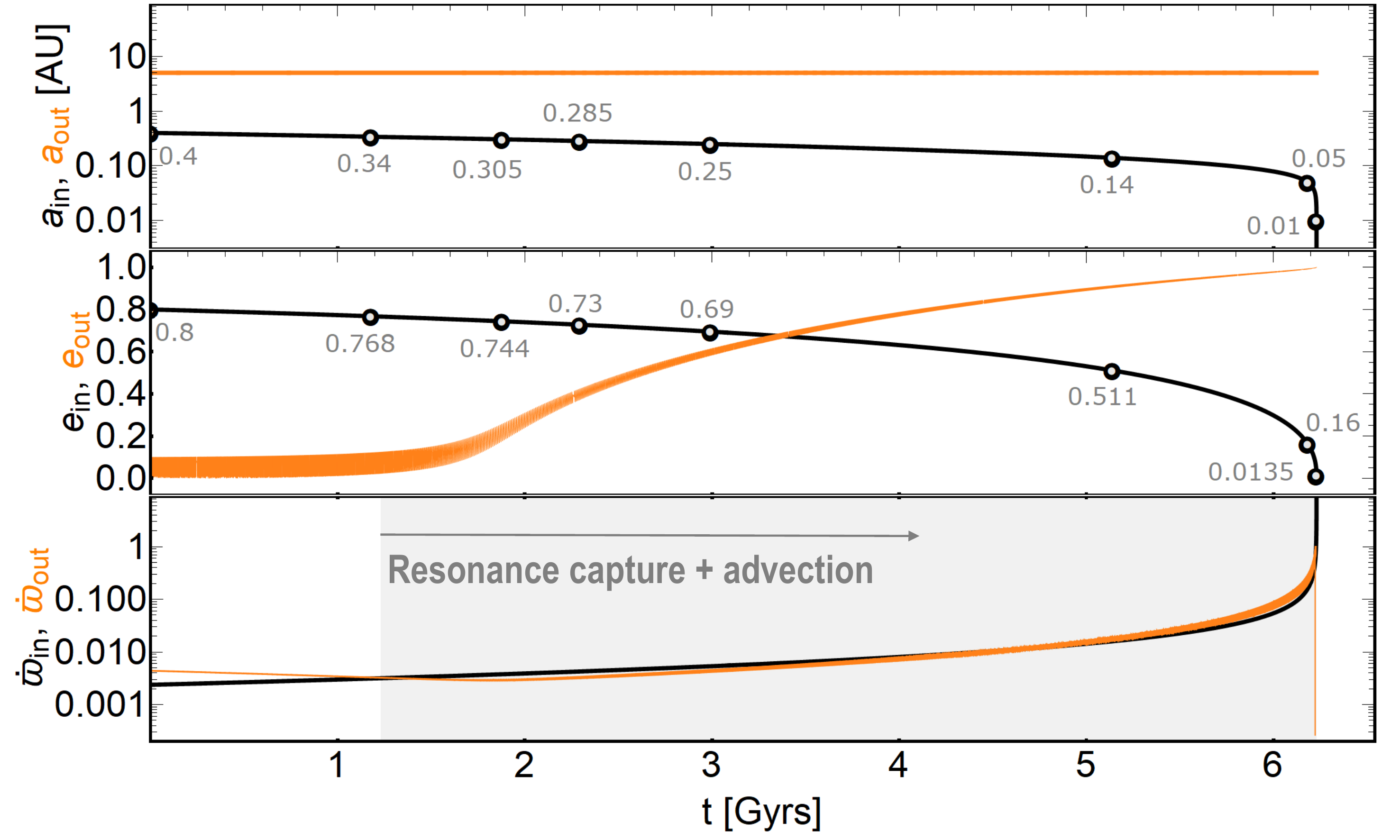}\\
\caption{The evolution of a merging BHB with a test-mass tertiary in a coplanar orbital configuration.
The masses of BHB are $m_1=46.15M_\odot$ and $m_2=13.85M_\odot$.
The initial semimajor axes and eccentricities of the inner and outer binaries are $a_{\IN,0}=0.4\au$, $e_{\IN,0}=0.8$,
$a_\OUT=5\au$ and $e_{\OUT,0}=0.1$, and the longitudes of the
periapsis of the two orbits are set to be $\varpi_{\IN,0}=\varpi_{\OUT,0}=0$.
The black (inner binary) and orange (outer binary) lines represent the numerical solutions of the DA secular equations
derived by the Hamiltonian (Eq. \ref{eq:Hamiltonian}).
Key moments during the inner BHB evolution are highlighted in the top two panels (gray dots) with the orbital parameter values labeled.
The bottom panel displays the apsidal precession rates given by Eqs. (\ref{eq: precession rate GR})-(\ref{eq: precession rate outer}).
}
\label{fig:Evolution}
\end{centering}
\end{figure}

To understand the dynamics of the outer binary influenced by the merging BHB,
we first consider the case where $m_3$ is a test-mass,
and thus the back-reaction from the tertiary on the inner binary can be ignored.

The top two panels of Fig. \ref{fig:Evolution} show an example of the orbital evolution of a coplanar triple.
The system satisfies the dynamical stability criterion \cite{Holman}.
The evolution is obtained by integrating the double-averaged
(DA; averaging over both the inner and outer orbital periods) secular equations of motion \cite{Liuetal-2015}.
All Newtonian effects (up to the octupole order), the leading order PN corrections, and
the dissipation due to gravitational wave (GW) are included in the calculations.
We see that the BHB undergoes orbital decay and eccentricity circularization due to GW radiation.
The merger time of the BHB is approximately given by
$T_\mathrm{merger}=[5c^5 a_{\IN,0}^4(1-e_{\IN,0}^2)^{7/2}]/[256 G^3 m_{12}^2 \mu_\IN]$ \cite{Peters-1964}.

\begin{figure*}
\begin{centering}
\includegraphics[width=17cm]{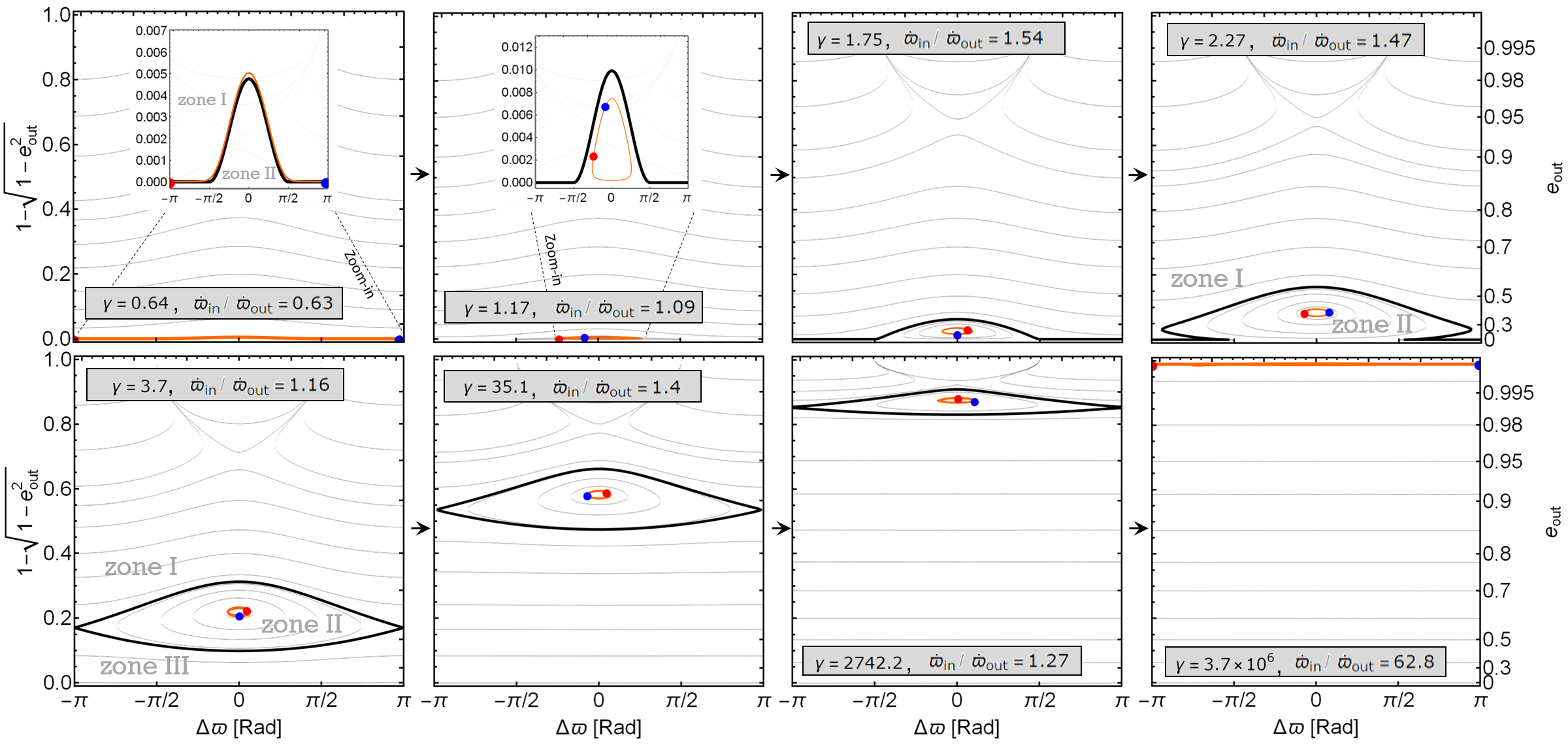}
\caption{Phase-space portraits in terms of the canonical variables ($1-\sqrt{1-e_\OUT^2}$ and $\Delta \varpi$)
at different stages of the orbital evolution for the system in our fiducial model (refer to Fig. \ref{fig:Evolution}).
In each panel, $\gamma\equiv\dot\varpi_\IN/\dot\varpi_\OUT|_{e_\OUT=0}$
and $\dot\varpi_\IN/\dot\varpi_\OUT$ are evaluated by
Eqs. (\ref{eq: CD}) and (\ref{eq: precession rate GR})-(\ref{eq: precession rate outer})
at the time labeled in Fig. \ref{fig:Evolution}.
The solid gray lines are the contours of constant $\hat{\mathcal{H}}$.
The existence of a separatrix (black line) divides phase-space into
circulation zones (zone I and III) and libration zone (zone II).
The orange curve represents the numerical solution of the outer binary's eccentricity vector $\bm e_\OUT$,
originating at the red dot and ending at the blue dot.
To better visualize both evolution trajectory and separatrix, zoom-in subfigures are included in the top-left panels.
}
\label{fig:phase space diagram}
\end{centering}
\end{figure*}

The bottom panel describes the time evolution of the apsidal precession rates,
in which $\dot\varpi_\IN$ is given by the 1st-order PN effect in the inner BHB
\be\label{eq: precession rate GR}
\dot\varpi_\IN=\frac{3G^{3/2}m_{12}^{3/2}}{c^2a_\IN^{5/2}(1-e_\IN^2)},
\ee
and $\dot\varpi_\OUT$ is due to the secular perturbation of the inner binary by the outer orbit \cite{Liuetal-2015,Liu-SMBHB}
\ba\label{eq: precession rate outer}
\dot\varpi_\OUT=\frac{3}{4}\bigg(\frac{G\mu_\IN m_3a_\IN^2}{a_\OUT^3}\bigg)\frac{1}{L_\OUT(1-e_\OUT^2)^{3/2}}.
\ea
We see that
starting from $\dot\varpi_\IN<\dot\varpi_\OUT$ initially,
$\dot\varpi_\IN$ increases and $\dot\varpi_\OUT$ decreases during orbital decay.
The outer eccentricity
can be excited once $\dot\varpi_\IN$ catches up to $\dot\varpi_\OUT$.
This resonance capture is precisely defined in the phase-space diagram (see Fig. \ref{fig:phase space diagram}).
Importantly, rather than a single resonance passage,
both frequencies evolve in lockstep until the inner BHB merges,
resulting in a significant growth of $e_\OUT$.
It's worth noting that the final $e_\OUT$ becomes quite large,
and the DA approximation may be challenged.
We employ alternative single-averaged equations of motion to assess its validity in the Supplemental material.

\textit{Theory.---}\label{sec 3}
To understand the nonlinear resonance capture and advection phenomenon
shown in Fig. \ref{fig:Evolution},
we consider the secular Hamiltonian of the system:
\ba\label{eq:Hamiltonian}
\mathcal{H}=-\frac{G m_1m_2}{2a_\IN}-\frac{G m_{12}m_3}{2a_\OUT}+\Phi_\mathrm{N}+\Phi_\mathrm{PN}.
\ea
Here $\Phi_\mathrm{N}$ is the Newtonian potential energy between the inner and outer orbits up to the octupole order, given by
\ba\label{eq:potential N}
\Phi_\mathrm{N}=\Phi\frac{-2-3e_\IN^2+\frac{15}{8}\varepsilon_{\oct} e_\IN (4+3e_\IN^2)\cos \Delta\varpi}
{6(1-e_\OUT^2)^{3/2}},
\ea
where
\ba
\Phi=\frac{3}{4}\frac{G\mu_\IN m_3 a_\IN^2}{a_\OUT^3},
\ea
and $\varepsilon_{\oct}=[(m_1-m_2)a_\IN e_\OUT]/[m_{12}a_\OUT(1-e_\OUT^2)]$ and
$\Delta\varpi\equiv\varpi_\OUT-\varpi_\IN$ with $\varpi_\IN$, $\varpi_\OUT$ being the longitudes of pericenters.
The term $\Phi_\mathrm{PN}$ is the PN potential that induces the apsidal precession of the inner binary:
\be\label{eq:potential GR}
\Phi_\mathrm{PN}=-\frac{3G^2m_1m_2m_{12}}{c^2 a_\IN^2\sqrt{1-e_\IN^2}}.
\ee
To examine the dynamics of the outer eccentricity vector $\bm{e}_\OUT$ in a frame corotating
with the inner eccentricity vector $\bm{e}_\IN$,
we can rewrite the Hamiltonian
as (after dropping the non-essential constant terms)
\ba\label{eq: ROT potential}
\mathcal{H}_\mathrm{rot}=\Phi_\mathrm{N}-\dot\varpi_\IN L_\OUT.
\ea
By introducing $\Phi_0=\Phi|_{a_\IN=a_{\IN,0}}$ with $a_{\IN,0}$ being the initial value of $a_\IN$,
Eq. (\ref{eq: ROT potential}) can be made dimensionless as
\ba\label{eq: dimensionless Hamiltonian}
\overline{\mathcal{H}}&&=\frac{\mathcal{H}_\mathrm{rot}}{\Phi_0}=A\hat{\mathcal{H}}\\
&&=A\bigg[\frac{B+C\frac{e_\OUT}{1-e_\OUT^2}\cos \Delta\varpi}{6(1-e_\OUT^2)^{3/2}}-\gamma\sqrt{1-e_\OUT^2}\bigg]\nonumber,
\ea
where the coefficients are defined as
\ba
&&A\equiv\bigg(\frac{a_\IN}{a_{\IN,0}}\bigg)^2,\label{eq: CA}\\
&&B\equiv-2-3e_\IN^2,\label{eq: CB}\\
&&C\equiv\frac{15}{8}\frac{m_1-m_2}{m_1+m_2}\frac{a_\IN}{a_\OUT} e_\IN (4+3e_\IN^2),\label{eq: CC}\\
&&\gamma\equiv\frac{\dot\varpi_\IN}{\dot\varpi'_\OUT}\simeq\frac{4Gm_{12}^3}{c^2m_1 m_2 a_\IN}
\bigg(\frac{a_\OUT}{a_{\IN}}\bigg)^{7/2}\frac{1}{1-e_\IN^2}\label{eq: CD}
\ea
with $\dot\varpi'_\OUT=\dot\varpi_\OUT(1-e_\OUT^2)^2$.
These coefficients evolve slowly as the BHB undergoes orbital decay driven by GW emission.

Fig. \ref{fig:phase space diagram} shows the dynamical behavior of the vector $\bm e_\OUT$ in the
phase-space at different stages of the BHB's orbital decay (marked by the value of $\gamma$).
Here, Poincar\'{e} variables are adopted, with $(1-\sqrt{1-e_\OUT^2})$ and $\Delta \varpi$ representing the canonical momentum and angle, respectively.
For each value of $\gamma$, the evolution of $\bm e_\OUT$ follows a level curve of constant $\hat{\mathcal{H}}$ (gray line),
either circulating freely from 0 to $2\pi$ or librating around a fixed $\Delta \varpi$ (i.e., resonance).
The circulating and librating trajectories of $\hat{\mathcal{H}}$ divide the phase-space into different zones,
bounded by the separatrix (thick black line) around a given $\gamma$,
the trajectory of $\bm e_\OUT$ (orange line) can only evolve in one of the zones.

In the beginning, $\gamma<1$ and $\bm e_\OUT$ circulates (top left panel of Fig. \ref{fig:phase space diagram}).
As the inner BHB decays, $\gamma$ increases. Around $\gamma\simeq1$,
the $\bm e_\OUT$ trajectory (orange line) crosses the separatrix,
and the circulating trajectory transitions into a librating one, indicating that the system is captured into the resonance
(the second panel of Fig. \ref{fig:phase space diagram}).
Although the total energy of the system is not conserved,
the evolution of $\bm e_\OUT$ can be described by a series of `frozen' Hamiltonians for each individual ($a_\IN$, $e_\IN$).
As $\gamma$ evolves in time, the separatrix gradually shifts to higher values of $e_\OUT$.
The librating trajectory is carried along with $\Delta \varpi\simeq0$ within zone II (termed `advection'),
accompanied by significant increase in $e_\OUT$.
During this `adiabatic' evolution within the separatrix, the phase-space area of the trajectory,
\ba\label{eq: phase space area}
\mathcal{A}=\oint \bigg(1-\sqrt{1-e_\OUT^2}\bigg)d\Delta \varpi,
\ea
is conserved.
Near the merger of the BHB, the area of zone II shrinks until the separatrix disappears, and the resonant trajectory transitions to circulating eventually
(see lower panels).

Note that apsidal precession resonance and the possibility of exciting eccentricity at resonance haven been well recognized in the planetary
dynamics literatures, often in the linear eccentricity regime (e.g., \cite{Heppenheimer,Ward,Lin,XuWenrui,Diego,Suyubo}).
However, the distinct advection effect,
sweeping through the full range of canonical momentum, is a unique feature of our present study.
Three requirements for the system to end up captured into resonance advection must be satisfied \cite{Quillen,Henrard}, including:

{\parindent0pt\textit{i) Adiabatic evolution}}.
The separatrix crossing must be slow.
Namely, the timescale of $e_\OUT-$excitation due to resonance should be shorter than the
merger time of BHB \cite{Liu-Daniel}, i.e., $\dot\varpi_\OUT T_\mathrm{merger}\gg1$, which is easily satisfied.

{\parindent0pt\textit{ii) Specific direction of time-varying $\gamma$}}.
The resonance encounter must be from
$\dot\varpi_\OUT>\dot\varpi_\IN$ towards $\dot\varpi_\OUT<\dot\varpi_\IN$.
For the opposite case, only a one-time jump in the eccentricity due to resonance crossing is excepted \cite{Liu-2020-PRD}.

{\parindent0pt\textit{iii) Sufficiently small initial eccentricity}}.
Resonance capture requires that the trajectory encounter the separatrix while zone III has not yet appeared.
This implies $e_{\OUT,0}\lesssim[15(m_1-m_2)a_\IN e_\IN (4+3e_\IN^2)]/(8m_{12}a_\OUT)$ (see Supplemtary material).

\begin{figure}
\begin{centering}
\includegraphics[width=8cm]{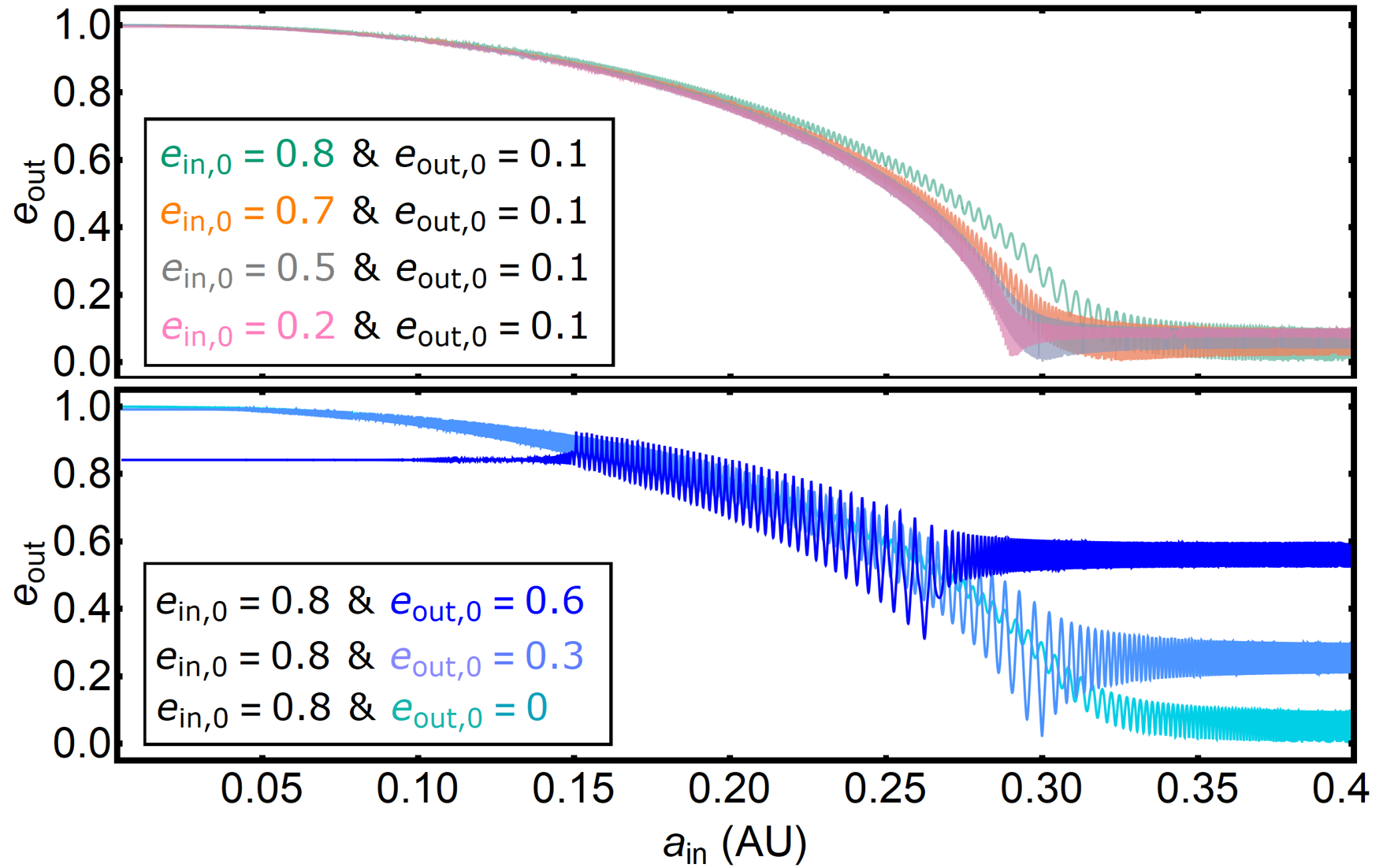}
\caption{The evolution of the outer binary eccentricity as a function of the inner binary semimajor axis.
We consider various combinations of ($e_{\IN,0}$, $e_{\OUT,0}$)
with the same parameters as in Fig. \ref{fig:Evolution}.
Note that for systems with $e_{\IN,0}=0.8$,
the maximum $e_{\OUT,0}$ allowed for a stable triple is $\sim0.66$ \cite{Holman}.
}
\label{fig:Initial eccentricity}
\end{centering}
\end{figure}

Fig. \ref{fig:Initial eccentricity} illustrates the excitation of stellar eccentricity for a range of $(e_{\IN,0}, e_{\OUT,0})$.
We find that the cases with $e_{\IN,0}\gtrsim e_{\OUT,0}$ always
give rise to resonant driving of $e_\OUT$, with weak dependence on the initial eccentricities.
This behavior can be understood as follows.
In the absence of separatrix crossing, as noted before,
the phase-space area $\mathcal{A}$ (Eq.\ref{eq: phase space area}) is an adiabatic invariant.
At the time of separatrix crossing,
$\mathcal{A}$ is not conserved, and the trajectory can transition into nearby zones \cite{Suyubo,Henrard}.
For a system initially set with a quasi-circular outer orbit,
the trajectory of $\bm e_\OUT$ circulates near the bottom part of the phase-space portrait (see the top left panel of Fig. \ref{fig:phase space diagram}).
Because there are only two zones at separatrix crossing,
the system is guaranteed to be captured into zone II
(as in the top panels of Fig. \ref{fig:phase space diagram}).
When $0.3\lesssim e_\OUT\lesssim0.6$, zone III emerges in the transition,
and both zone II and III become available.
The probabilities of transition to zone $\rm II$ and $\rm III$ are given by
\cite{Henrard-1982,Henrard-1987}
\be
P_{\rm I\rightarrow \rm II}=-\frac{\partial\mathcal{A}_{\rm II}/\partial\gamma}{\partial\mathcal{A}_{\rm I}/\partial\gamma}
=-\frac{\partial\mathcal{A}_{\rm II}}{\partial\mathcal{A}_{\rm I}},\label{eq: PII}
\ee
\be
P_{\rm I\rightarrow \rm III}=-\frac{\partial\mathcal{A}_{\rm III}/\partial\gamma}{\partial\mathcal{A}_{\rm I}/\partial\gamma}
=-\frac{\partial\mathcal{A}_{\rm III}}{\partial\mathcal{A}_{\rm I}}\label{eq: PIII},
\ee
where $\mathcal{A}_{\rm I(\rm II,III)}$ is the area of the individual zone.
Note that $d\mathcal{A}_{\rm I}+d\mathcal{A}_{\rm II}+d\mathcal{A}_{\rm III}=0$.
In our case, both $P_{\rm I\rightarrow \rm II}$ and $P_{\rm I\rightarrow \rm III}$ are positive
due to the expansion of zone II and zone III and the contraction of zone I at the separatrix crossing.
Whether the system transitions to zone II or zone III depends on the initial $\Delta \varpi_0$,
which affects the initial phase-space area for a given $e_{\OUT,0}$.
Note that when $e_{\OUT,0}=0.6$, the resonant eccentricity excitation terminates prematurely, and $e_\OUT$ reaches around $0.8$.
This occurs because after resonance capture,
the phase-space area enclosed by the trajectory becomes too large, making separatrix crossing inevitable as zone II begins to shrink
(see Supplemental material).
If $e_{\OUT,0}$ is even higher,
capture into zone II becomes impossible as $P_{\rm I\rightarrow \rm II}$ is negative at the time of separatrix crossing.

\begin{figure}
\begin{centering}
\includegraphics[width=7cm]{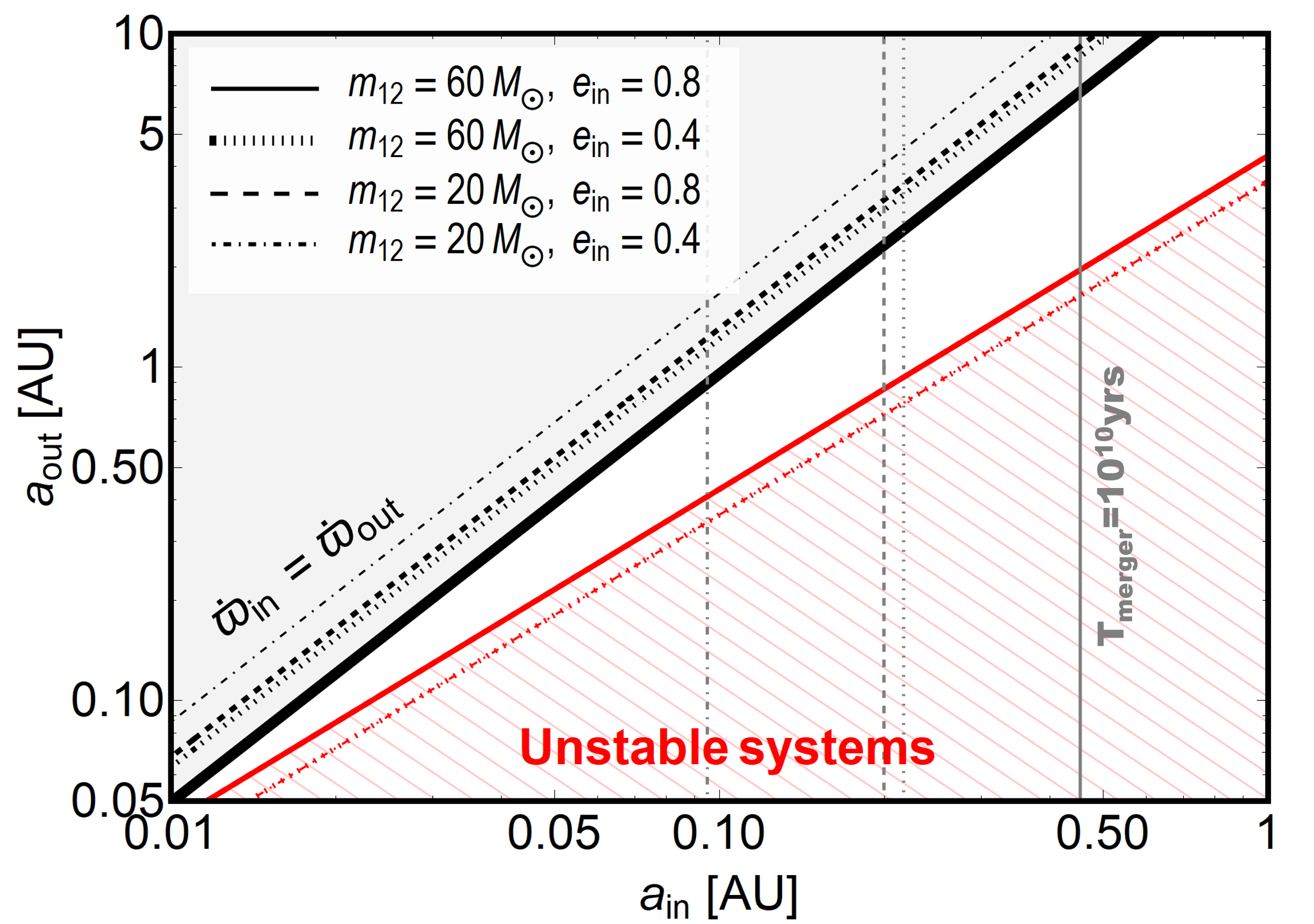}
\caption{Parameter space in the $a_\IN$-$a_\OUT$ plane showing which system may undergo resonance capture and advection
during the orbital decay of the inner BHB.
We consider different types of BHBs as labeled.
Assuming the tertiary body is a test mass with a circular orbit,
the cross-hatched region indicates dynamically unstable systems,
and the gray-shaded region signifies systems with $\gamma\gtrsim1$, or $\dot\varpi_\IN\gtrsim\dot\varpi_\OUT$.
The vertical lines refer to the merger time $T_\mathrm{merger}=10^{10}$yrs.
}
\label{fig:parameter space}
\end{centering}
\end{figure}

As mentioned earlier, a necessary condition for the resonance capture is $\gamma\lesssim 1$ for the initial system.
This translate into the constraint that the initial $a_\OUT$ should be larger than a few times of the instability limit \cite{Holman}
(see Fig. \ref{fig:parameter space}).

\textit{Finite $m_3$: modified evolution of merging BHBs.---}
We now consider the `BHB$+$star' system where the tertiary companion has a finite mass $m_3=1M_\odot$.
In this case, the perturbation from the tertiary companion on the inner binary cannot be ignored.
Since the outer binary is allowed to `gain' some eccentricity from the inner BHB at resonance,
the inner binary may experience a decrease in eccentricity due to the transfer of angular momentum,
therefore reducing the orbital decay rate of the BHB.
When the initial $e_\OUT$ is large, the inner binary can `gain' some eccentricity, therefore increasing the decay rate.

\begin{figure}
\begin{centering}
\includegraphics[width=8cm]{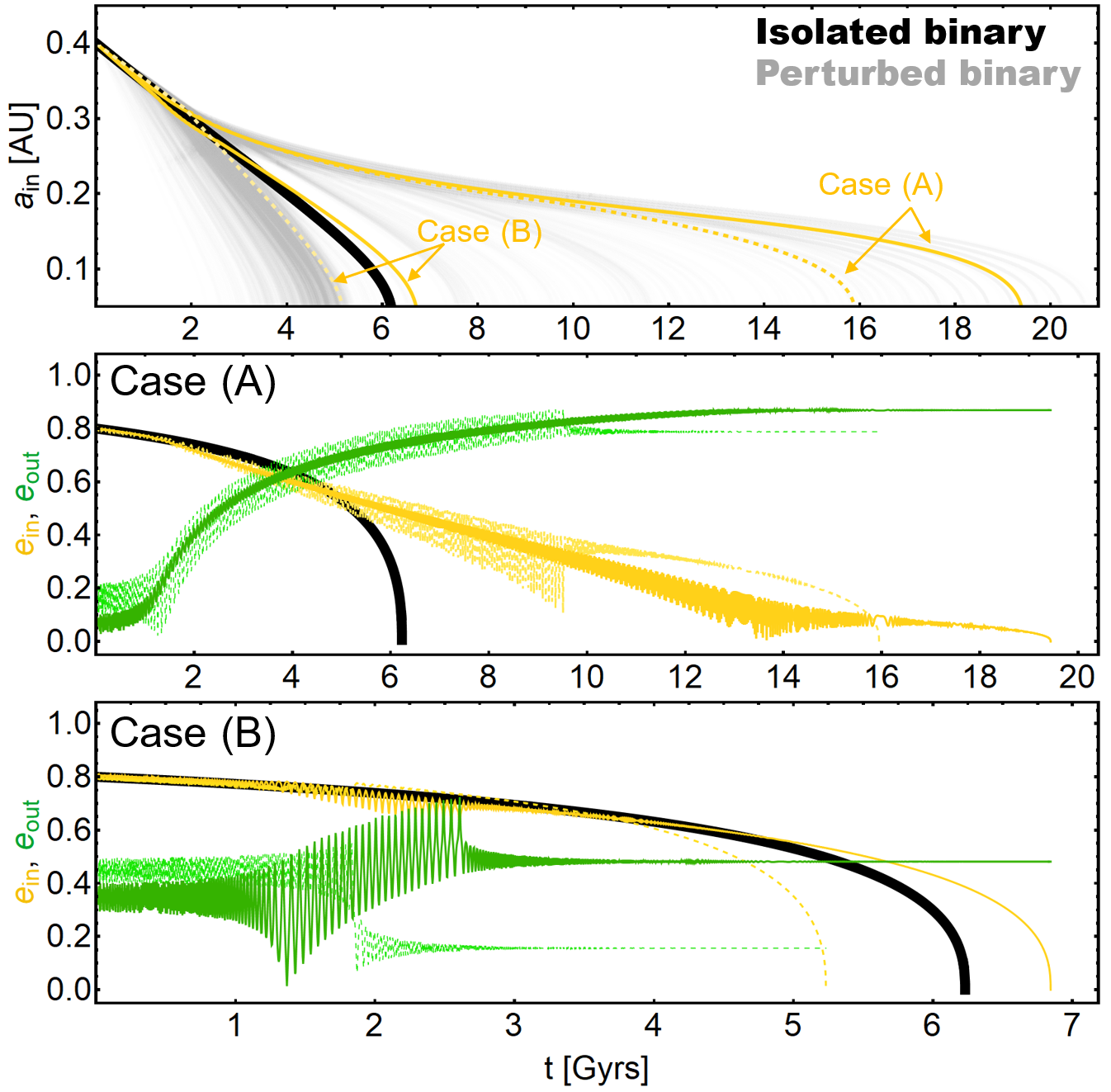}
\caption{Top panel: the evolution of merging BHBs with and without the perturbation of a tertiary star.
We use our fiducial model (see Fig. \ref{fig:Evolution} for the parameters) with various $a_\OUT$, $e_{\OUT,0}$
(constrained by Fig. \ref{fig:parameter space}) and $\Delta\varpi_0$ values.
Lower panels: the evolution of orbital eccentricities for the system that have undergone resonance passage or capture.
We choose $a_\OUT=5.3\au$ and $e_{\OUT,0}=0.1$ for case (A) and $a_\OUT=5.4\au$ and $e_{\OUT,0}=0.4$ for case (B).
In each panel, the solid and dashed lines correspond to systems initialized with $\Delta\varpi_0=0$ and $\pi$, respectively.
}
\label{fig:BHB star}
\end{centering}
\end{figure}

To explore how the evolution of the BHB can be altered, we
begin with the BHB with $m_1=46.15M_\odot$, $m_2=13.85M_\odot$, $a_{\IN,0}=0.4\au$ and $e_{\IN,0}=0.8$ in our fiducial model.
Based on Fig. \ref{fig:parameter space},
we consider the tertiary star with $a_\OUT$ in the range of $(5, 10)\au$ and all possible eccentricities within the stability limit.
Random initial $\Delta\varpi_0$ values are chosen.
The results are presented in panel (A) of Fig. \ref{fig:BHB star}.
We see that the BHB may experience either slower or faster hardening compared to the isolated binary.
The BHB with a longer merger time experiences extra eccentricity decrease
due to the tertiary companion, associated with the eccentricity excitation of the tertiary.
On the other hand, the BHB with the shorter merger time gains eccentricity from the outer companion,
accompanied by decrease in $e_\OUT$.

The middle and lower panels of Fig. \ref{fig:BHB star} depict four examples of the eccentricity evolution
for different $e_{\OUT,0}$ and $\Delta\varpi_0$ values,
corresponding to different initial phase-space trajectories of the outer binary.
For $e_{\OUT,0}=0.1$ (middle panel),
we see that $e_\OUT$ experiences resonant excitation and eventually freezes at a finite value;
at the same time, $e_\IN$ experiences `extra' decrease during the resonance advection,
leading to a decelerated merger event.
For $e_{\OUT,0}=0.4$ (lower panel),
the eccentricity evolution is sensitive to $\Delta\varpi_0$
(or the initial phase-space trajectory); the
merger undergoes either advection-deceleration or acceleration,
as a result of the probabilistic transition during the separatrix crossing.

\textit{Summary and discussion.---}
We have shown that a star orbiting around an inspiraling BHB can experience
a distinct resonance capture and advection phenomenon,
leading to a significant increase in the eccentricity.
This resonance occurs when the apsidal precession rates of the inner and outer orbits are comparable.
The resonance advection gives rise to the apsidal alignment,
enabling a continuous exchange of angular momentum between the two orbits until the BHB merges.
The eccentricity growth requires the inner binary to have an unequal masses and an elliptical orbit,
and the outer orbit to have a relatively small eccentricity.
For systems with initially large outer eccentricities,
the resonance capture occurs probabilistically or may not occur at all.
In this study, we have restricted to coplanar configurations.
Our preliminary study shows that as long as the mutual inclinations between the inner and outer orbits is small ($\lesssim30^\circ$),
similar resonance phenomenon can happen.

The dynamical behavior of the outer binary eccentricity has several implications.
The exchange of angular momentum between the inner  and outer binaries alters the evolution of
$e_\IN$, thereby influencing the hardening rate of the BHBs.
Additionally, the notable increase in $e_\OUT$ serves as a distinctive signature for
revealing hidden BHBs \cite{Liu-Daniel} that could be observed by
up-coming GW detectors like LISA \cite{LISA}, TianQin \cite{TianQin}, Taiji \cite{TaiJi} and B-DECIGO \cite{DECIGO}.
In some cases, significant $e_\OUT-$excitation could lead to a small pericenter distance for the outer stellar orbit,
resulting significant tidal dissipation or even disruption of the star \cite{MTDE}.

The occurrence rate of the `BHB$+$star' systems studied in this \textit{letter} is unknown.
The progenitor stars of these BHBs often expand to hundreds or thousands of solar radii
during their evolution, potentially interacting dynamically with the stellar tertiary.
However, low-metallicity massive stellar binaries could remain compact \cite{Mandel-2016,Marchant-2016,duBuisson,Riley},
allowing for dynamically-stable compact triples \cite{Alejandro2021}.
Alternatively, a BHB might initially form and subsequently capture a long-lived low-mass star.

The resonant eccentricity excitation mechanism studied here
can be applied to other astrophysical systems,
including a planet orbiting a binary star and
a stellar-mass object around a supermassive BHB with or without a gaseous disc.
We leave these topics to the future investigations.

\textit{Acknowledgement.---}
B.L. thanks Yubo Su for useful discussions.
B.L. gratefully acknowledges support from
the European Union's Horizon 2021 research and innovation programme under the Marie Sklodowska-Curie grant agreement No. 101065374.

\onecolumngrid
\appendix

\newpage
\clearpage
\setcounter{equation}{0}
\setcounter{figure}{0}
\setcounter{table}{0}
\setcounter{page}{1}
\makeatletter

\begin{center}
\textbf{\large Supplemental Material\\[0.5ex]
Extreme resonant eccentricity excitation of stars around merging black hole binary
}
\end{center}

\section{Equations of motion}
\label{sec:A}

\begin{figure*}
\begin{centering}
\includegraphics[width=8.3cm]{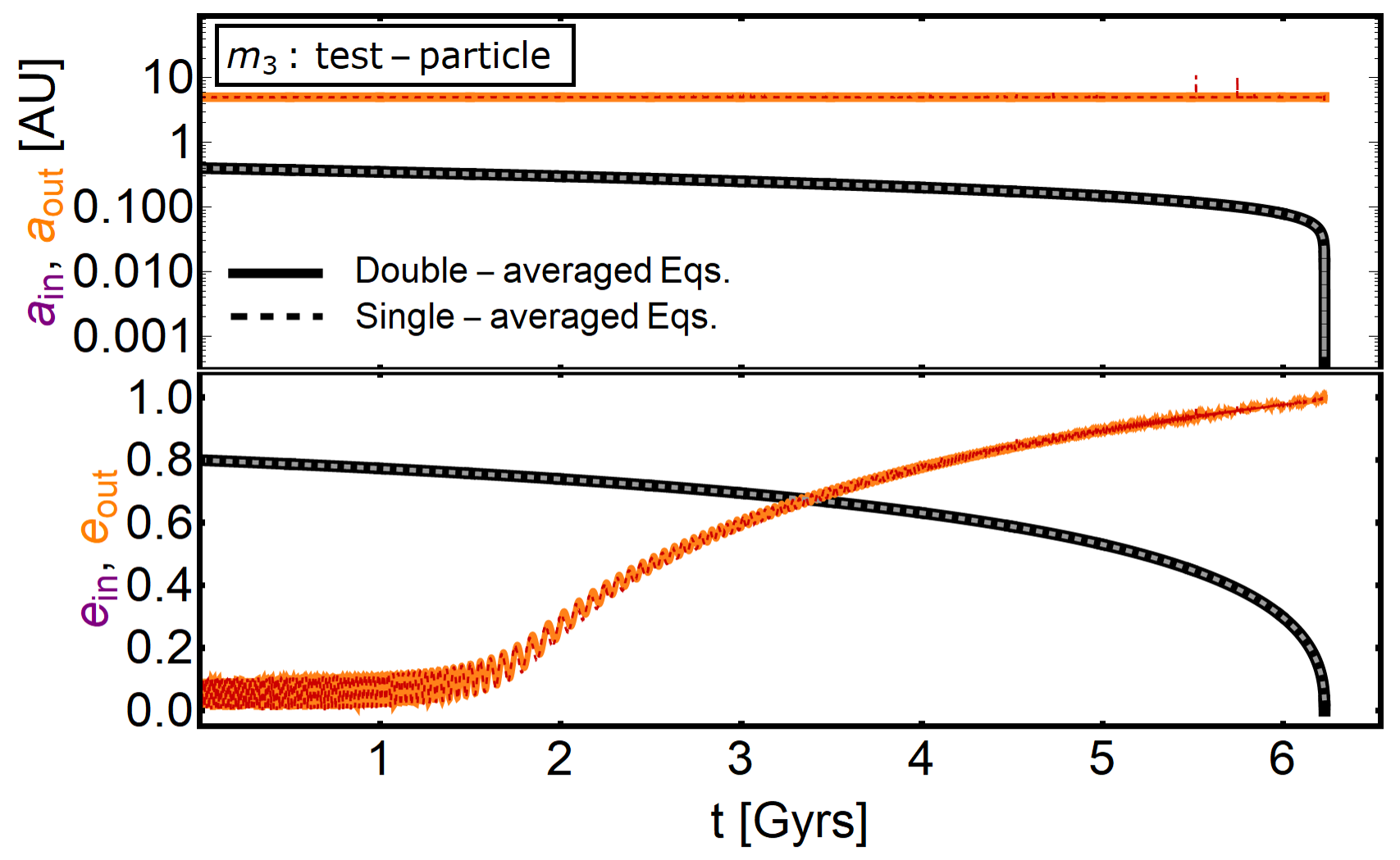}
\includegraphics[width=8.3cm]{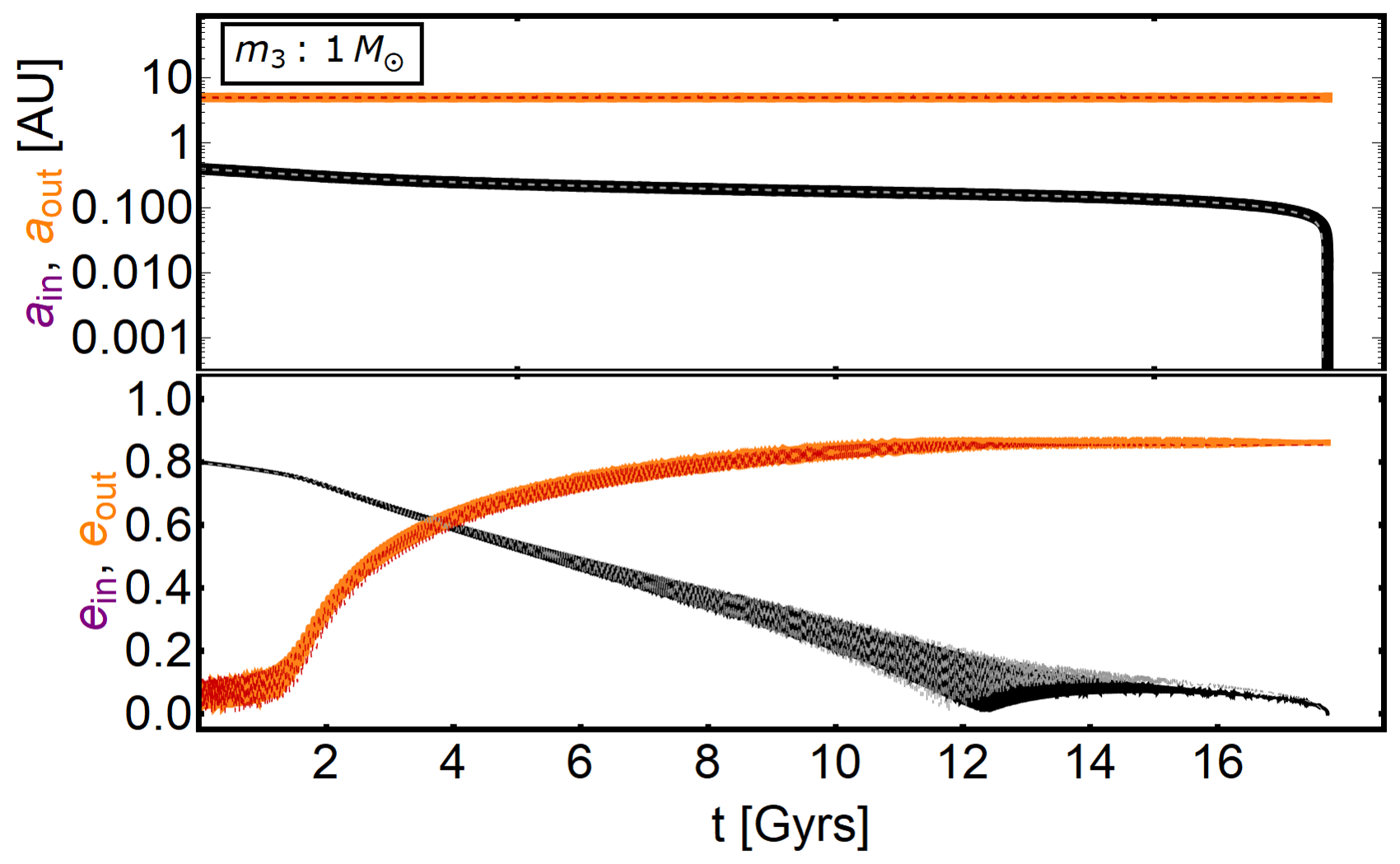}
\caption{Sample orbital evolution of a BHB with a tertiary companion.
System parameters are the same as our fiducial example from Fig. 1 in the main text, except for the mass of the tertiary companion (as labeled).
The solid and dashed lines denote two integration methods.
In the SA calculations, the initial true anomaly of the outer binary is set to be $\pi$.
}
\label{fig:DASA evolution}
\end{centering}
\end{figure*}

The secular interaction timescale of the outer orbit driven by the inner binary is $|\dot\varpi_\OUT|^{-1}$ (Eq. 3),
where
\ba
\dot\varpi_\OUT\sim\frac{\mu_\IN}{m_{12}}\bigg(\frac{a_\IN}{a_\OUT}\bigg)^2(1-e_\OUT^2)^{-2}n_\OUT.
\ea
On the other hand, the pericenter frequency of the outer orbit is
\ba
n_p\sim n_\OUT (1-e_\OUT^2)^{-3/2}.
\ea
The double-averaged secular equations are valid when $n_p\gg\dot\varpi_\OUT$, or
$\frac{\mu_\IN}{m_{12}}\big(\frac{a_\IN}{a_\OUT}\big)^2(1-e_\OUT^2)^{-1/2}\ll1$.

To calibrate our different approaches, Fig. \ref{fig:DASA evolution} presents
the examples of the orbital evolution of a BHB with a coplanar companion,
obtained by using double-averaged (DA) and single-averaged (SA) secular equations \cite{Liuetal-2015,Liu-ApJ}.
The outcomes obtained from these two integration methods exhibit near-identical results.
The evolution achieved by the SA equations permits exchanges in both energy and angular momentum.
We find weak oscillations in the evolution of $a_\OUT$ in the upper panel
when $e_\OUT$ attains a sufficiently large value (depicted by the dashed line).
These oscillations can be attributed to the stability of the system.

\section{Resonance capture and advection}
\label{sec:B}

\begin{figure*}
\begin{centering}
\includegraphics[width=18cm]{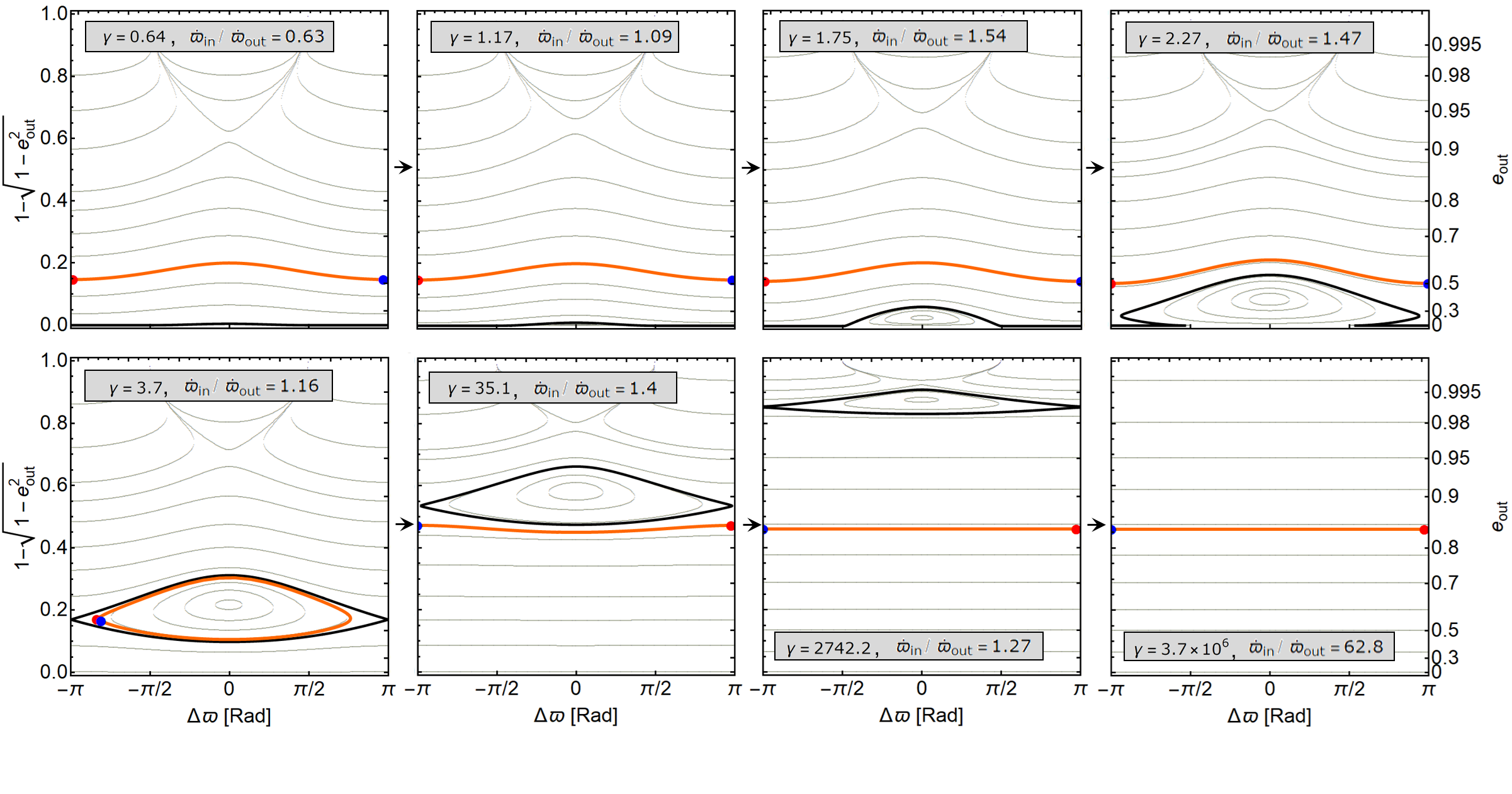}\\
\hspace{1pt}
\includegraphics[width=18cm]{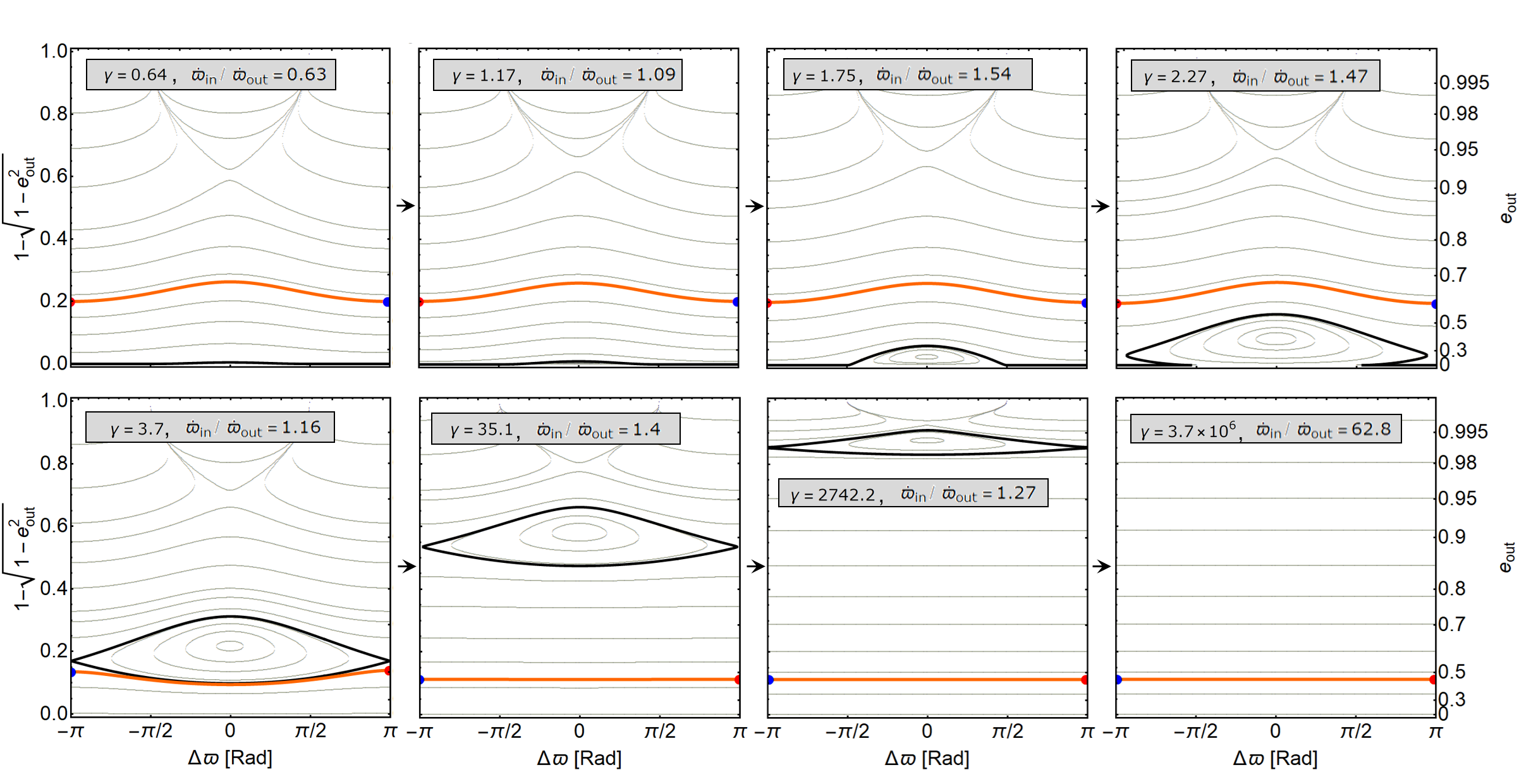}
\caption{Similar to Fig. 2 in the main text, but we set the initial eccentricity to $e_{\OUT,0}=0.6$
with $\Delta\varpi_0=0$ (top two panels) and $\Delta\varpi_0=\pi$ (bottom two panels).
}
\label{fig:phase space diagram 2}
\end{centering}
\end{figure*}

\begin{figure*}
\begin{centering}
\includegraphics[width=18cm]{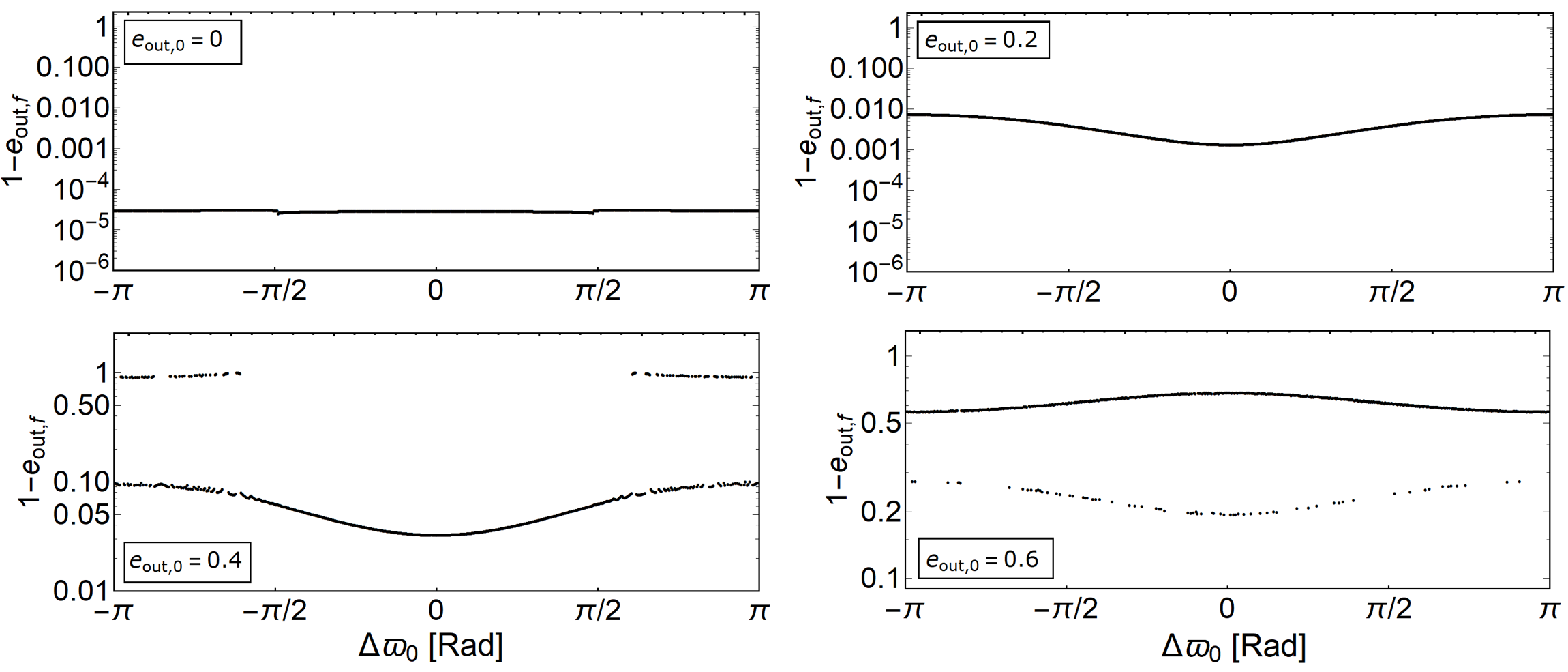}
\caption{The final $e_\OUT$ as a function of $\Delta\varpi_0\equiv\varpi_{\OUT,0}-\varpi_{\IN,0}$
for different values of $e_{\OUT,0}$.
The system studied here is from our fiducial model, in which
the outer binary is set to $a_\OUT=5\au$ with the initial eccentricity being specified as labeled.
}
\label{fig:eout window}
\end{centering}
\end{figure*}

In this section, we explore the dynamics of the outer binary influenced by the merging inner BHB,
initially set with non-zero $e_\OUT$.
We again focus on our fiducial model with $m_1=46.15M_\odot$, $m_2=13.85M_\odot$, $a_{\IN,0}=0.4\au$ and $e_{\IN,0}=0.8$,
and consider the tertiary companion is a test-mass.

Similar to Fig. 2 in the main text, Fig. \ref{fig:phase space diagram 2}
depicts the dynamic evolution of the vector $\bm e_\OUT$ with an initial value of $e_{\OUT,0}=0.6$.
As there is no back-reaction from the tertiary on the inner binary,
the structure of the phase-space portrait remains the same as Fig. 2,
featuring the same contours of constant $\hat{\mathcal{H}}$ and the separatrix for each $\gamma$.
However, in contrast to the case with $e_{\OUT,0}\ll1$,
the trajectory of $\bm e_\OUT$ here encounters the separatrix around $\gamma\sim3$
for which zone III already exists.
With zone II expanding, both the transition probabilities into zone II and zone III
($P_{\rm I\rightarrow \rm II}$ and $P_{\rm I\rightarrow \rm III}$)
are positive (Eqs. 14-15).
We find that system initialized with different $\Delta\varpi_0$ can undergo either resonance capture or passage,
as shown in the upper and lower panels of Fig. \ref{fig:phase space diagram 2}.
The dependence arises because, for a given $e_{\OUT,0}$,
the enclosed phase-space areas of $\mathcal{A}_{\rm I(\rm II,III)}$ are different for different $\Delta\varpi_0$.
Note that for the case of $\Delta\varpi_0=0$,
the librating evolution trajectory transitions to circulating at $\gamma\simeq35.1$.
This is because zone II starts shrinking in area at this moment,
while the phase-space area covered by the evolution trajectory is still conserved.
The separatrix crossing becomes unavoidable, and the resonant eccentricity excitation terminates prematurely.

Fig. \ref{fig:eout window} examines the impact of $e_{\OUT,0}$ and $\Delta\varpi_0$ for the resonance passage and advection.
For each $e_{\OUT,0}$, the final value of outer binary eccentricity
is computed as a function of $\Delta\varpi_0$.
We see that for the systems with $e_{\OUT,0}\ll1$,
the resonance advection is guaranteed.
When $e_{\OUT,0}\gtrsim0.4$, the distribution of $e_{\OUT,\mathrm{f}}$ imply that
resonance advection occurs probabilistically.

\section{Apsidal resonance: theoretical analysis}
\label{sec:C}

As mentioned in the main text, the dimensionless secular Hamiltonian of a coplanar triple is given by
\ba\label{eq: dimensionless Hamiltonian 2}
\hat{\mathcal{H}}=\frac{1}{6(1-e_\OUT^2)^{3/2}}\bigg[B+C\frac{e_\OUT}{1-e_\OUT^2}\cos \Delta\varpi\bigg]-\gamma\sqrt{1-e_\OUT^2},
\ea
where the coefficients are defined as
\ba
&&B=-|B|=-(2+3e_\IN^2),\\
&&C=\frac{15}{8}\frac{m_1-m_2}{m_1+m_2}\frac{a_\IN}{a_\OUT} e_\IN (4+3e_\IN^2),\\
&&\gamma=\frac{4Gm_{12}^3}{c^2m_1 m_2 a_\IN}
\bigg(\frac{a_\OUT}{a_{\IN}}\bigg)^{7/2}\frac{1}{1-e_\IN^2}.
\ea
By assuming $e_\OUT^2\ll1$, $\hat{\mathcal{H}}$ (Eq. \ref{eq: dimensionless Hamiltonian 2})
can be further simplified to (dropping a non-essential constant and a multiplicity factor)
\be\label{eq: dimensionless Hamiltonian 3}
\hat{\mathcal{H}}=-De_\OUT^2+Ce_\OUT\cos \Delta\varpi
\ee
with
\be
D=\frac{3}{2}|B|-3\gamma.
\ee
During the orbital decay of the inner BHB, the coefficients $|B|$ (of order a few) and $C$ (typically $\ll1$) decrease in time,
while $\gamma$ starts from $\lesssim1$ and increases to large value.
For the system studied in this \textit{letter}, $D>0$ is valid at the early stage of the orbital decay,
the evolution of the outer binary eccentricity is thus determined by
\be\label{eq: eout solution}
e_\OUT=\frac{1}{2}\bigg[\beta\cos \Delta\varpi\pm\sqrt{\beta^2\cos^2\Delta\varpi+4\hat{\mathcal{H}}'}\bigg],
\ee
where $\beta = C/D$ and $\hat{\mathcal{H}}'=\hat{\mathcal{H}}/D$.

\begin{figure*}
\begin{centering}
\includegraphics[width=8cm]{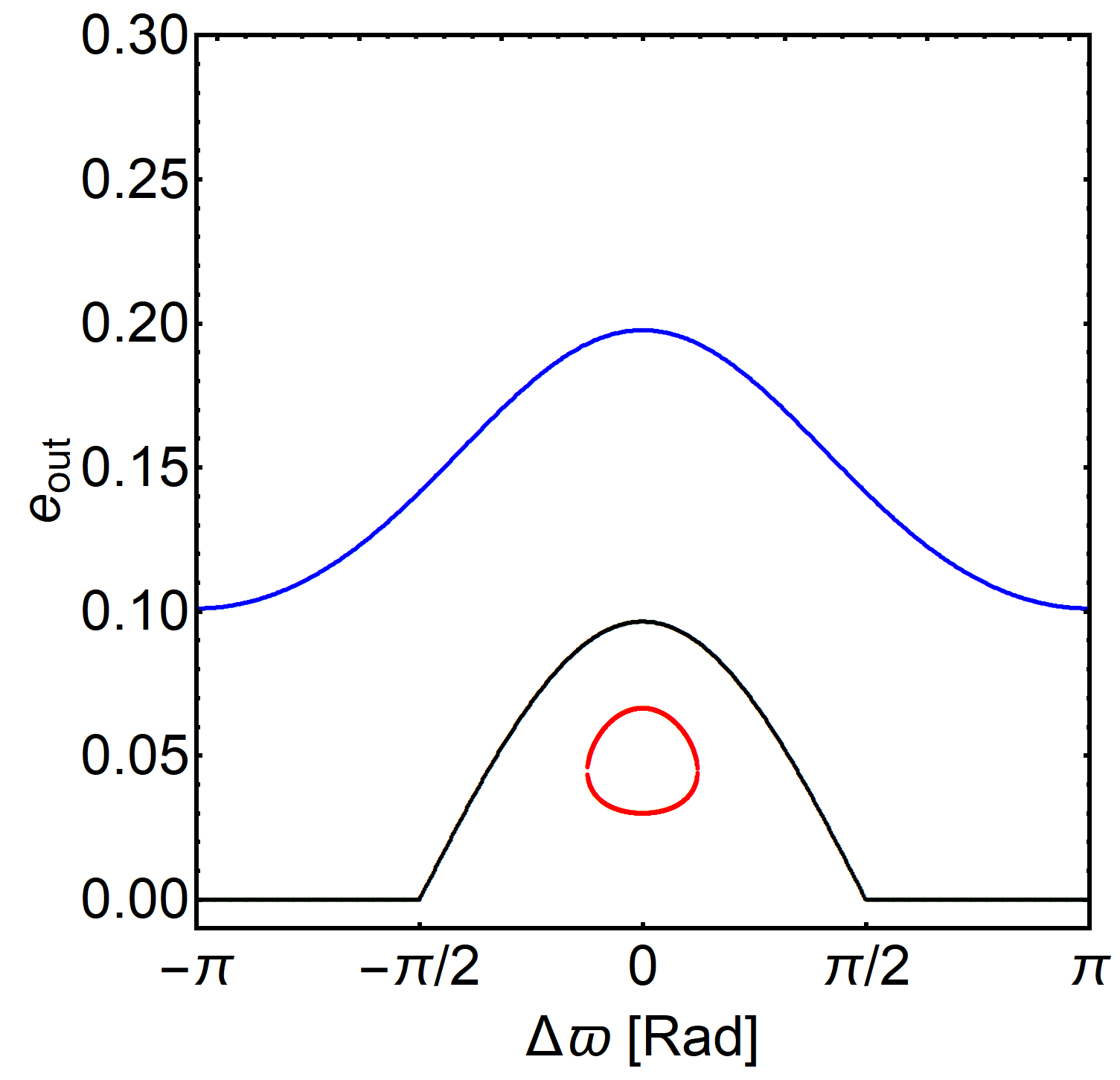}
\caption{Phase-space portraits in terms of $e_\OUT$ and $\Delta \varpi$.
Three lines are given by Eq. (\ref{eq: eout solution}) when $\hat{\mathcal{H}}>0$ (blue), $\hat{\mathcal{H}}=0$ (black) and $\hat{\mathcal{H}}<0$ (red).
Here, $\beta$ is evaluated by the initial orbital parameters shown in Fig. 1 in the main text.
}
\label{fig:phase space diagram 3}
\end{centering}
\end{figure*}

Fig. \ref{fig:phase space diagram 3} shows the solutions of Eq. (\ref{eq: eout solution}).
Similar to  Fig. 2 in the main text, we see that a separatrix (black line; corresponding to $\hat{\mathcal{H}}=0$) devides the phase-space into two zones,
above which the outer binary eccentricity follows the circulating trajectory (blue line; as $\hat{\mathcal{H}}>0$) and
below which $e_\OUT$ is librating (red line; as $\hat{\mathcal{H}}<0$).
Importantly, the peak value of $e_\OUT$ of the separatrix is given by $e_{\OUT, \mathrm{peak}}=\beta$, which occurs at $\Delta \varpi=0$.

Therefore, to have secure resonance capture, the requirements can be summarized as:\\
\textit{i)} Initially, $D$ should be greater than zero, which translates to $\gamma\lesssim1$;\\
\textit{ii)} Adiabatic evolution.\\
\textit{iii)} Specific direction of time-varying $\gamma$.\\
\textit{iv)} Sufficiently small initial eccentricity. Based on the theoretical analysis above,
the requirement for $e_{\OUT,0}$ can now be well quantified as
\ba
e_{\OUT,0}&&\lesssim e_{\OUT, \mathrm{peak}}~(\mathrm{at~~\gamma\sim1})\\
&&\simeq\frac{15}{8}\frac{m_1-m_2}{m_1+m_2}\frac{a_\IN}{a_\OUT} e_\IN (4+3e_\IN^2)~(\mathrm{at~~\gamma\sim1})\nonumber.
\ea


\begin{thebibliography}{100}

\bibitem{LIGO-2021} R. Abbott et al. (LIGO Scientific, VIRGO), arXiv:2108.01045 [gr-qc] (2021).

\bibitem{Lipunov-1997} V. M. Lipunov, K. A. Postnov, and M. E. Prokhorov, Astron. Lett. {\bf 23}, 492 (1997).

\bibitem{Lipunov-2007} V. M. Lipunov, V. Kornilov, E. Gorbovskoy, D. A. H. Buckley, N. Tiurina, P. Balanutsa, A. Kuznetsov, J. Greiner,
V. Vladimirov, D. Vlasenko et al., Mon. Not. R. Astron. Soc. {\bf 465}, 3656 (2017).

\bibitem{Podsiadlowski-2003} P. Podsiadlowski, S. Rappaport, and Z. Han, Mon. Not. R. Astron. Soc. {\bf 341}, 385 (2003).

\bibitem{Belczynski-2010} K. Belczynski, M. Dominik, T. Bulik, R. O'Shaughnessy,
C. Fryer, and D. E. Holz, Astrophys. J. Lett. {\bf 715}, L138 (2010).

\bibitem{Belczynski-2016} K. Belczynski, D. E. Holz, T. Bulik, and R. O'Shaughnessy, Nature (London) {\bf 534}, 512 (2016).

\bibitem{Dominik-2012} M. Dominik, K. Belczynski, C. Fryer, D. E. Holz, E. Berti, T. Bulik, I. Mandel, and R. O'Shaughnessy,
Astrophys. J. {\bf 759}, 52 (2012).

\bibitem{Dominik-2013} M. Dominik, K. Belczynski, C. Fryer, D. E. Holz, E. Berti, T. Bulik, I. Mandel, and R. O'Shaughnessy,
Astrophys. J. {\bf 779}, 72 (2013).

\bibitem{Dominik-2015} M. Dominik, E. Berti, R. O'Shaughnessy, I. Mandel, K. Belczynski, C. Fryer, D. E. Holz, T. Bulik, and F. Pannarale,
Astrophys. J. {\bf 806}, 263 (2015).

\bibitem{Alejandro-2017} S. Stevenson, A. Vigna-G\'omez, I. Mandel, J. W. Barrett, C. J. Neijssel, D. Perkins, and
S. E. de Mink, Nature Commun. {\bf 8}, 14906 (2017).

\bibitem{Mandel-2016} I. Mandel and S. E. De Mink, Mon. Not. R. Astron. Soc. {\bf 458}, 2634 (2016).

\bibitem{Marchant-2016} P. Marchant, N. Langer, P. Podsiadlowski, T. M. Tauris, and T. J. Moriya, Astron. Astrophys. {\bf 588}, A50 (2016).

\bibitem{duBuisson} L. du Buisson, P. Marchant, P. Podsiadlowski, C. Kobayashi, F. B. Abdalla, P. Taylor, I. Mandel,
S. E. de Mink, T. J. Moriya, and N. Langer, Mon. Not. R. Astron. Soc. {\bf 499}, 5941 (2020).

\bibitem{Riley} J. Riley, I. Mandel, P. Marchant, E. Butler, K. Nathaniel, C. Neijssel, Shortt S., and A. Vigna-G\'omez,
Mon. Not. R. Astron. Soc. {\bf 505}, 663 (2021).

\bibitem{Baruteau-2011} C. Baruteau, J. Cuadra, and D. N. C. Lin, Astrophys. J. {\bf 726}, 28 (2011).

\bibitem{McKernan-2012} B. McKernan, K. E. S. Ford, W. Lyra, and H. B. Perets, Mon. Not. Roy. Astron. Soc. {\bf 425}, 460 (2012).

\bibitem{McKernan-2018} B. McKernan, K. E. S. Ford, J. Bellovary, N. W. C. Leigh, Z. Haiman, B. Kocsis, W. Lyra, M. M. MacLow, B. Metzger, M. O'Dowd,
S. Endlich, and D. J. Rosen, Astrophys. J. {\bf 866}, 66 (2018).

\bibitem{Bartos-2017} I. Bartos, B. Kocsis, Z. Haiman, and S. M{\'a}rka, Astrophys. J. {\bf 835}, 165 (2017).

\bibitem{Stone-2017} N. C. Stone, B. D. Metzger, and Z. Haiman, Mon. Not. Roy. Astron. Soc. {\bf 464}, 946 (2017).

\bibitem{Leigh-2018} N. W. C. Leigh, A. M. Geller, B. McKernan, K. E. S. Ford, M. M. Mac Low, J. Bellovary, Z. Haiman, W. Lyra,
J. Samsing, M. O’Dowd, B. Kocsis, and S. Endlich,  Mon. Not. Roy. Astron. Soc. {\bf 474}, 5672 (2018).

\bibitem{Secunda-2019} A. Secunda, J. Bellovary, M.-M. Mac Low, K. E. S. Ford, B. McKernan, N. W. C. Leigh, W. Lyra, and Z. S{\'a}ndor,
Astrophys. J. {\bf 878}, 85 (2019).

\bibitem{Yang-2019} Y. Yang, I. Bartos, V. Gayathri, K. E. S. Ford, Z. Haiman, S. Klimenko, B. Kocsis, S. M{\'a}rka, Z. M{\'a}rka,
B. McKernan, and R. O'Shaughnessy, Phys. Rev. Lett. {\bf 123}, 181101 (2019).

\bibitem{Grobner-2020} M. Gr\"{o}bner, W. Ishibashi, S. Tiwari, M. Haney, and P. Jetzer, Astron. Astrophys. {\bf 638}, A119 (2020).

\bibitem{Ishibashi-2020}W. Ishibashi and M. Gr\"{o}bner, Astron. Astrophys. {\bf 639}, A108 (2020).

\bibitem{Tagawa-2020} H. Tagawa, Z. Haiman, and B. Kocsis, Astrophys. J. {\bf 898}, 25 (2020).

\bibitem{Liyaping-2021} Y. P. Li, A.~M. Dempsey, S. Li, H. Li, and J. Li, Astrophys. J. {\bf 911}, 124 (2021).

\bibitem{Ford-2021} K. E. S. Ford and B. McKernan, arXiv:2109.03212 [astro-ph.HE] (2021).

\bibitem{Samsing-Nature} J. Samsing, I. Bartos, D.~J. D'Orazio, Z. Haiman, B. Kocsis, N.~W.~C. Leigh, B. Liu, M.~E. Pessah, and H. Tagawa,
Nature, {\bf 603}, 237 (2022).

\bibitem{Lirixin-2022} R. Li and D. Lai, arXiv:2202.07633 [astro-ph.HE] (2022).

\bibitem{Lijiaru-2022} J. Li and D. Lai, arXiv:2203.05584 [astro-ph.HE] (2022).

\bibitem{Zwart(2000)} S. F. P. Zwart and S. L. W. McMillan, Astrophys. J. Lett. {\bf 528}, L17 (2000).

\bibitem{OLeary(2006)} R. M. O'Leary, F. A. Rasio, J. M. Fregeau, N. Ivanova, and R. OShaughnessy, Astrophys. J. {\bf 637}, 937 (2006).

\bibitem{Miller(2009)} M. C. Miller and V. Lauburg, Astrophys. J. {\bf 692}, 917 (2009).

\bibitem{Banerjee(2010)} S. Banerjee, H. Baumgardt, and P. Kroupa, Mon. Not. R. Astron. Soc. {\bf 402}, 371 (2010).

\bibitem{Downing(2010)} J. M. B. Downing, M. J. Benacquista, M. Giersz, and R. Spurzem, Mon. Not. R. Astron. Soc. {\bf 407}, 1946 (2010).

\bibitem{Ziosi(2014)} B. M. Ziosi, M. Mapelli, M. Branchesi, and G. Tormen, Mon. Not. R. Astron. Soc. {\bf 441}, 3703 (2014).

\bibitem{Rodriguez(2015)} C. L. Rodriguez, M. Morscher, B. Pattabiraman, S. Chatterjee, C.-J. Haster, and F. A. Rasio,
Phys. Rev. Lett. {\bf 115}, 051101 (2015).

\bibitem{Samsing(2017)} J. Samsing and E. Ramirez-Ruiz, Astrophys. J. Lett. {\bf 840}, L14 (2017).

\bibitem{Samsing(2018)} J. Samsing and D. J. D'Orazio, Mon. Not. R. Astron. Soc. {\bf 481}, 5445 (2018).

\bibitem{Rodriguez(2018)} C. L. Rodriguez, P. Amaro-Seoane, S. Chatterjee, and F. A. Rasio, Phys. Rev. Lett. {\bf 120}, 151101 (2018).

\bibitem{Gondan(2018)} L. Gond\'{a}n, B. Kocsis, P. Raffai, and Z. Frei, Astrophys. J. {\bf 860}, 5 (2018).

\bibitem{vonZeipel} H. von Zeipel, Astronomische Nachrichten {\bf 183}, 345 (1910).

\bibitem{Lidov} M. L. Lidov, Planetary and Space Science {\bf 9}, 719 (1962).

\bibitem{Kozai} Y. Kozai, Astron. J. {\bf 67}, 591 (1962).

\bibitem{Smadar} S. Naoz, Annu. Rev. Astron. Astrophys. {\bf 54}, 441 (2016).

\bibitem{Miller-2002} M. C. Miller and D. P. Hamilton, Astrophys. J. {\bf 576}, 894 (2002).

\bibitem{Wen-2003} L. Wen, Astrophys. J. {\bf 598}, 419 (2003).

\bibitem{Antonini-2012} F. Antonini and H. B. Perets, Astrophys. J. {\bf 757}, 27 (2012).

\bibitem{Antonini(2017)} F. Antonini, S. Toonen, and A. S. Hamers, Astrophys. J. {\bf 841}, 77 (2017).

\bibitem{Silsbee(2017)} K. Silsbee and S. Tremaine, Astrophys. J. {\bf 836}, 39 (2017).

\bibitem{Petrovich-2017} C. Petrovich and F. Antonini, Astrophys. J. {\bf 846}, 146 (2017).

\bibitem{Liu-ApJ} B. Liu and D. Lai, Astrophys. J. {\bf 863}, 68 (2018).

\bibitem{Xianyu-2018} L. Randall and Z.-Z. Xianyu, Astrophys. J. {\bf 853}, 93 (2018).

\bibitem{Hoang-2018} B.-M. Hoang, S. Naoz, B. Kocsis, F. A. Rasio, and F. Dosopoulou, Astrophys. J. {\bf 856}, 140 (2018).

\bibitem{Liu-Quadruple} B. Liu and D. Lai, Mon. Not. R. Astron. Soc. {\bf 483}, 4060 (2019).

\bibitem{Fragione-Quadruple} G. Fragione and B. Kocsis, Mon. Not. R. Astron. Soc. {\bf 486}, 4781 (2019).

\bibitem{Fragione-nulearcluster} G. Fragione, E. Grishin, N. W. C. Leigh, H. B. Perets, and R. Perna, Mon. Not. R. Astron. Soc. {\bf 488}, 47 (2019).

\bibitem{Zevin-2019} M. Zevin, J. Samsing, C. Rodriguez, C.-J. Haster, and E. Ramirez-Ruiz, Astrophys. J. {\bf 871}, 91 (2019).

\bibitem{Liu-HierarchicalMerger} B. Liu and D. Lai, Mon. Not. R. Astron. Soc. {\bf 502}, 2049 (2021).

\bibitem{Michaely-2019} E. Michaely and H. B. Perets, Astrophys. J. Lett. {\bf 887}, L36 (2019).

\bibitem{Michaely-2020}E. Michaely and H. B. Perets, Mon. Not. R. Astron. Soc. {\bf 498}, 4924 (2020).

\bibitem{Liu-Yuan} B. Liu, D. Lai, and Y.-F. Yuan, Phys. Rev. D {\bf 92}, 124048 (2015).

\bibitem{Liu-2020-PRD} B. Liu, and D. Lai, Phys. Rev. D {\bf 102}, 023020 (2020).

\bibitem{Moe-2017}M. Moe and R. Di Stefano, ApJS {\bf 230}, 15 (2017).

\bibitem{Tokovinin} A. Tokovinin, S. Thomas, M. Sterzik, and S. Udry, Astron. Astrophys. {\bf 450}, 681 (2006).

\bibitem{Raghavan} D. Raghavan, H. A. McAlister, T. J. Henry, D. W. Latham, G. W. Marcy, B. D. Mason, D. R. Gies, R. J. White,
and T. A. ten Brummelaar, Astrophys. J. Suppl. Ser. {\bf 190}, 1 (2010).

\bibitem{Fuhrmann} K. Fuhrmann, R. Chini, L. Kaderhandt, and Z. Chen, Astrophys. J. {\bf 836}, 139 (2017).

\bibitem{Ransom} S. M. Ransom, I. H. Stairs, A. M. Archibald, J. W. T. Hessels, D. L. Kaplan, M. H. van Kerkwijk, J. Boyles, A. T. Deller,
S. Chatterjee, and A. Schechtman-Rook et al., Nature {\bf 505}, 520 (2014).

\bibitem{Thompson} T. A. Thompson, C. S. Kochanek, K. Z. Stanek, C. Badenes, R. S. Post, T. Jayasinghe, D. W. Latham, A. Bieryla, G. A. Esquerdo,
P. Berlind, M. L. Calkins, J. Tayar, L. Lindegren, J. A. Johnson, T. W. S. Holoien, K. Auchettl, and K. Covey, Science {\bf 366}, 637 (2019)

\bibitem{Eisner} N.~L. Eisner, C. Johnston, S. Toonen, A.~J. Frost, S. Janssens, C.~J. Lintott, S. Aigrain, et al.,
Mon. Not. R. Astron. Soc. {\bf 511}, 4710 (2022).

\bibitem{Suto-1} T. Hayashi, S. Wang, and Y. Suto, Astrophys. J. {\bf 890}, 112 (2020).

\bibitem{Suto-2} T. Hayashi and Y. Suto ,Astrophys. J. {\bf 897}, 29 (2020).

\bibitem{Suto-3} T. Hayashi and Y. Suto, Astrophys. J. {\bf 907}, 48 (2021).

\bibitem{Liu-Daniel}B. Liu, D.~J. D'Orazio , A. Vigna-G{\'o}mez, J. Samsing, Phys. Rev. D {\bf 106}, 123010 (2022).

\bibitem{Holman} M.~J. Holman and P.~A. Wiegert,  Astron. J. {\bf 117}, 621 (1999).

\bibitem{Liuetal-2015} B. Liu, D.~J. Mu{\~n}oz, and D. Lai, Mon. Not. Roy. Astron. Soc. {\bf 447}, 747 (2015).

\bibitem{Peters-1964} P. C. Peters, Phys. Rev. {\bf 136}, B1224 (1964).

\bibitem{Liu-SMBHB}B. Liu, D. Lai, Mon. Not. R. Astron. Soc. {\bf 513}, 4657 (2022).

\bibitem{Heppenheimer} T.~A. Heppenheimer, Icarus {\bf 41}, 76 (1980).

\bibitem{Ward} W.~R. Ward, Icarus {\bf 47}, 234 (1981).

\bibitem{Lin} M. Nagasawa, D.~N.~C. Lin, S. Ida, Astrophys. J. {\bf 586}, 1374 (2003).

\bibitem{XuWenrui}W. Xu, D. Lai,  Mon. Not. R. Astron. Soc. {\bf 468}, 3223 (2017).

\bibitem{Diego} D.~J. Mu{\~n}oz, N.~C. Stone, C. Petrovich, F.~A. Rasio, 2022, arXiv, arXiv:2204.06002.

\bibitem{Suyubo} Y. Su, D. Lai, Astrophys. J. {\bf 903}, 7 (2020).

\bibitem{Quillen} A.~C. Quillen, Mon. Not. R. Astron. Soc. {\bf 365}, 1367 (2006).

\bibitem{Henrard} J. Henrard, The Adiabatic Invariant in Classical Mechanics (Berlin, Heidelberg: Springer), 117–235 (1993).

\bibitem{Henrard-1982} J. Henrard, Celestial mechanics {\bf 27}, 1 (1982).

\bibitem{Henrard-1987} J. Henrard and C. Murigande, Celestial mechanics {\bf 40}, 345 (1987).

\bibitem{LISA} P. Amaro-Seoane, H. Audley, S. Babak, J. Baker, E. Barausse, P. Bender, E. Berti, P. Binetruy et al.,
arXiv:1702.00786 [astro-ph.IM] (2017).

\bibitem{TianQin} J. Luo, L.-S. Chen, H.-Z. Duan, Y.-G. Gong, S. Hu, J. Ji, Q. Liu, J. Mei, V. Milyukov, M. Sazhin, C.-G. Shao, V. T.
Toth, H.-B. Tu, Y. Wang, Y. Wang, H.-C. Yeh, M.-S. Zhan, Y. Zhang, V. Zharov, and Z.-B. Zhou, Classical and Quantum Gravity {\bf 33}, 035010 (2016).

\bibitem{TaiJi} W.-R. Hu and Y.-L. Wu, National Science Review {\bf 4}, 685 (2017).

\bibitem{DECIGO} T. Nakamura, M. Ando, T. Kinugawa, H. Nakano, K. Eda, S. Sato, M. Musha, T. Akutsu, T. Tanaka,
N. Seto, N. Kanda, and Y. Itoh, Progress of Theoretical and Experimental Physics {\bf 2016}, 093E01 (2016).

\bibitem{MTDE} H.~B. Perets, Z. Li, J.~C. Lombardi, S.~R. Milcarek, Astrophys. J. {\bf 823}, 113 (2016).

\bibitem{Alejandro2021} A. Vigna-G\'omez, S. Toonen, E. Ramirez-Ruiz, N. W. C.
Leigh, J. Riley, and C.-J. Haster, Astrophys. J. Lett. {\bf 907}, L19 (2021).

\end{thebibliography}
\end{document}